\documentstyle[prl,aps,preprint]{revtex}
\begin{document}
\draft
\title{Correlation functions in the three-chain Hubbard ladder}
\author{Takashi Kimura, Kazuhiko Kuroki and Hideo Aoki}
\address{Department of Physics, University of Tokyo, 
Hongo, Tokyo 113, Japan}
\date{\today}
\maketitle
\begin{abstract}
In order to check whether 
odd-numbered Tomonaga-Luttinger ladders are dominated by 
antiferromagnetic correlations associated with gapless spin 
excitations, correlation functions of 
the doped three-chain Hubbard model are 
obtained with the bosonization 
at the renormalization-group fixed point. 
The correlation of the singlet superconducting pairing 
across the central and edge chains is found to be dominant, 
reflecting two gapful spin modes, while the intra-edge spin density wave 
correlation, reflecting the gapless mode, is only subdominant. 
This implies that, when there are multiple spin modes, 
a dominant superconductivity can arise from the presence of 
{\it some} spin gap(s) despite the coexistence of power-law 
correlated spins. 
\end{abstract} 
\pacs{74.20.Mn and 71.27.+a}
Recently, a wealth of experimental and theoretical 
results indicate that interacting electrons on multiple chains, 
or ladders, are an interesting realm of correlated systems.  
An increasing fascination toward them has been kicked off 
by an `even-odd' conjecture by Rice et al., 
who have proposed that 
the ladder, at half-filling, with 
even number of chains should be 
a spin liquid reflecting 
the absence of gapless spin excitations, while 
odd-numbered chains should be 
antiferromagnetic (AF) reflecting 
the presence of gapless spin excitations.
\cite{Rice,White,Greven}
This is reminiscent of Haldane's conjecture
\cite{Haldane,Affleck}
for the one-dimensional (1D) 
AF Heisenberg model for 
integer and half-odd-integer spins.  

When the system is doped with carriers, it is usually supposed that 
an even-numbered ladder should 
exhibit the interchain 
singlet superconductivity as expected from 
the persistent spin gap, while an odd-numbered ladder 
should have the usual $2k_F$ spin-density wave (SDW) reflecting 
the gapless spin excitations. 
In 1D, an interacting electron system may be described 
by the Tomonaga-Luttinger liquid. 
\cite{Lutt}
Thus intensive analytical studies have been performed 
by extending the Tomonaga-Luttinger model analysis 
to the two-chain ladders. 
These analytical calculations support the superconductivity in 
double chains within the perturbational renormalization-group 
analysis for weak repulsive interactions.
\cite{Fabrizio,Finkelstein,Schulz,Balents,Nagaosa}  
To be more precise, the correlation of the interchain pairing 
is dominant and much stronger than that of the subdominant $4k_F$ 
charge density wave (CDW) in the above calculations.
On the other hand, numerical calculations performed directly 
for the two-chain $t-J$ and Hubbard models have also been performed,
although the phase diagram including the strong-coupling regime 
has not been conclusive.  
\cite{Dagotto,Tsunetsugu,Noack,Hayward,Sano}

Experimentally, cuprates  
SrCu$_2$O$_3$ and Sr$_2$Cu$_3$O$_5$ are investigated as 
prototypes of two and three chain systems, respectively\cite{kitaoka}.
The two-chain system indeed shows a spin liquid behavior 
characteristic of a finite spin-correlation length, 
while the three-chain system 
shows an AF behavior. 

Theoretically, however, whether the 
`even-odd' conjecture continues to be valid for triple chains 
remains an open question.  
In fact, Arrigoni has looked into the triple chains having 
weak interactions by using 
the usual perturbational renormalization-group technique 
to conclude that gapless and gapful spin excitations 
{\it coexist} there.\cite{Arri} 
Namely, he has actually enumerated the numbers of gapless 
charge and spin modes on the phase diagram spanned by 
the doping level and the interchain electron tunneling strength.  
He found that, at half-filling, one gapless spin mode exists for 
the interchain hopping comparable with the intrachain hopping, 
in agreement with some experimental 
results and theoretical expectations. 
Away from the half-filling, on the other hand, 
one gapless spin mode is found to remain at the fixed point 
in the region where the fermi level intersects all the 
three bands in the noninteracting case. 
>From this, Arrigoni argues that 
the spin-spin correlation should decay as a power law. 

On the other hand, 
his result also indicates that two gapful spin modes 
exist in addition.  
While the existence of a gapful spin mode 
crudely favors a singlet superconductivity (SS), 
we are in fact faced here with an intriguing problem of what happens 
when gapless and gapful spin modes coexist, 
since it may well be possible that the presence of gap(s) in 
{\it some} out of multiple 
spin modes may be sufficient for a dominance of superconductivity. 
This has motivated us, in the present work, to actually 
look at the correlation functions 
using the bosonization method at the fixed point away from 
half-filling.  
Since we have the cuprate ladder in mind, we concentrate on 
the open boundary condition (OBC) across the chain, where 
the central chain is inequivalent to the two edge chains.  
We find that the interchain SS pairing 
between the central and edge chains is the dominant 
correlation, which is indeed realized due to 
the presence of the two gapful spin modes. 
On the other hand, the SDW correlation, 
which has a power law for the intra-edge chain 
reflecting the gapless spin mode, 
is only subdominant. 
                                                        
We start from the Hamiltonian, 
\begin{eqnarray}
H&=&H_0+H_{\rm int}, \\
H_0&=&\sum_{irk\sigma}\epsilon_k a^{\dagger}_{irk\sigma}a_{irk\sigma}
   \nonumber\\&&-t\sum_{rk\sigma}
(a^{\dagger}_{\alpha rk\sigma}a_{\beta rk\sigma}
+a^{\dagger}_{\beta rk\sigma}a_{\gamma rk\sigma}+{\rm h.c.}).
\end{eqnarray}
Here $a^{\dagger}_{irk\sigma}$ 
creates an electron 
with lattice momentum $k$ and spin $\sigma$ on right ($r=R$)
or left ($r=L$) going branch in the $i$-th chain 
($i=\alpha,\beta,\gamma$ with 
$\beta$ being the central one), 
$\epsilon_k$ 
the kinetic energy of each chain, 
and $t$ the interchain hopping. 
The one-electron part, $H_0$, may be diagonalized by a 
linear transformation,
\begin{eqnarray}
\left( \begin{array}{c} 
a_{\alpha rk\sigma} \\ a_{\beta rk\sigma} \\ a_{\gamma rk\sigma}
\end{array} \right)
=
\left( \begin{array}{ccc}
\frac{1}{2} & \frac{\sqrt{2}}{2} & \frac{1}{2}\\
\frac{1}{\sqrt{2}} & 0 & -\frac{1}{\sqrt{2}}\\
\frac{1}{2} & -\frac{\sqrt{2}}{2} & \frac{1}{2}
\end{array} \right)
\left( \begin{array}{c}
c_{1rk\sigma} \\ c_{2rk\sigma} \\ c_{3rk\sigma}
\end{array} \right)
\end{eqnarray}
resulting in 
\begin{eqnarray}
H_0&=&\sum_{rk\sigma}[(\epsilon_k-\sqrt{2} t)
c^{\dagger}_{1rk\sigma}c_{1rk\sigma}+
\epsilon_k c^{\dagger}_{2rk\sigma}c_{2rk\sigma}\nonumber\\
&&+(\epsilon_k + \sqrt{2} t)c^{\dagger}_{3rk\sigma}c_{3rk\sigma}].
\end{eqnarray}
Hereafter we linearize the band structure around the fermi points
as usual and neglect the difference in the fermi velocities 
of three bands, which will be acceptable for the weak-hopping case\cite{vf}.
We focus on the case in which all of three bands
are away from half-filling. 

The part of the Hamiltonian, $H_{\rm d}$, that can be 
diagonalized in the bosonization only includes 
forward-scattering processes in the band picture, 
and has the form
\begin{eqnarray}
H_{\rm d}&=&H_{\rm spin}+H_{\rm charge}, \nonumber\\
H_{\rm spin}&=&\sum_i\frac{v_{\sigma i}}{4\pi}\int\ dx
[\frac{1}{K_{\sigma i}}(\partial_x\phi_{i+})^2+
K_{\sigma i}(\partial_x\phi_{i-})^2], \\
H_{\rm charge}&=&\sum_i\frac{v_{\rho i}}{4\pi}\int\ dx
[\frac{1}{K_{\rho i}}(\partial_x\chi_{i+})^2
+K_{\rho i}(\partial_x\chi_{i-})^2].\nonumber
\end{eqnarray}
Here $\phi_{i+}$ is the spin phase field 
of the $i$-th band, $\chi_{i+}$ is the diagonal charge phase field, 
while 
$\phi_{i-}$($\chi_{i-}$) is the field dual to $\phi_{i+}$($\chi_{i+}$), 
$K_{\sigma i}$($K_{\rho i}$) the correlation exponent 
for the $\phi$($\chi_i$) phase 
with $v_{\sigma i}$($v_{\rho i}$) being their velocities. 
For the Hubbard type interaction, we 
have $v_{\sigma i}=v_F$, $K_{\sigma i}$=1 for all $i$'s, while 
$v_{\rho 1}=v_F$, $v_{\rho 2}=v_F\sqrt{1-4g^2}$, 
$v_{\rho 3}=v_F\sqrt{1-g^2/4}$, $K_{\rho 1}=1$, 
$K_{\rho 2}=\sqrt{(1-2g)/(1+2g)}$,
$K_{\rho 3}=\sqrt{(1-g/2)/(1+g/2)}$, where $g=U/2\pi v_F$ 
is the Hubbard $U$ interaction made dimensionless. 

The diagonalized charge field $\chi_{i\pm}$ is linearly 
related to the initial charge field $\theta_{i\pm}$
of the $i$-th band as
\begin{eqnarray}
\left( \begin{array}{c}
\theta_{1\pm}\\ \theta_{2\pm}\\ \theta_{3\pm}
\end{array} \right)
=
\left( \begin{array}{ccc}
\frac{1}{\sqrt{2}} & \frac{1}{\sqrt{3}} & \frac{1}{\sqrt{6}}\\
                 0 & \frac{1}{\sqrt{3}} & -\sqrt{\frac{2}{3}}\\
-\frac{1}{\sqrt{2}} & \frac{1}{\sqrt{3}} & \frac{1}{\sqrt{6}}
\end{array} \right)
\left( \begin{array}{c}
\chi_{1\pm}\\ \chi_{2\pm}\\ \chi_{3\pm}
\end{array} \right),
\end{eqnarray}
where $\theta_{i\pm}$ and $\phi_{\pm}$ are related to the field 
operator for electrons $\psi_{ir\sigma}$ as
\begin{eqnarray}
\psi_{i+(-)\sigma}(x)&=&\frac{\eta_{i+(-)\sigma}}{2 \pi \Lambda}
{\rm exp}\{ \pm ik_{iF}x\nonumber\\ &&\pm \frac{i}{2}
\{ \theta_{i+}(x) \pm \theta_{i-}(x)+(\phi_{i+}(x)\pm\phi_{i-}(x))]\}.
\end{eqnarray}
Here $\eta_{ir\sigma}$'s are majorana fermion operators
(or Haldane's $U$ operators)
\cite{U}
which ensure the anti-commutation relations between electron 
operators 
through the relation, $\{\eta_{ir\sigma},\eta_{i'r'\sigma'}\}_+=
2\delta_{ii'}\delta_{rr'}\delta_{\sigma\sigma'}$, 
$\eta_{ir\sigma}^{\dagger}=\eta_{ir\sigma}$.

There are still many scattering processes
corresponding to the backward scattering and pair tunneling scattering 
processes between two bands, which cannot be treated exactly. 
Arrigoni examined the effect of such scattering 
processes by diagrammatic perturbational renormalization group technique.
He found that the backward-scattering interactions within 
the first or the third band turn from positive to negative 
as the renormalization is performed 
and that the pair tunneling processes between the first 
and third bands also become relevant. 
At the fixed point the Hamiltonian density, $H^{*}$, 
then takes the form, in term of the phase variables, 
\begin{eqnarray}
H^{*}&=&-\frac{g_{b}(1)}{\pi^2 \Lambda^2}{\rm cos}(2\phi_{1+}(x))
            -\frac{g_{b}(3)}{\pi^2 \Lambda^2}{\rm cos}(2\phi_{3+}(x))
\nonumber\\
          &&+\frac{2g_{ft}(1,3)}{\pi^2 \Lambda^2}
            {\rm cos}(\sqrt{2}\chi_{1-}(x))
            {\rm sin}\phi_{1+}(x){\rm sin}\phi_{3+}(x),
\end{eqnarray}
where $g_{b}(1)$, $g_{b}(3)$ are negative large quantities 
and $g_{ft}(1,3)$ is a positive large quantity\cite{majo}. 

This indicates that the phase fields 
$\phi_{1+}$, $\phi_{3+}$, and $\chi_{1-}$ are
long-range ordered and fixed 
at $\pi/2$, $\pi/2$, and $\pi/\sqrt{2}$, respectively, 
which in turn implies that 
the correlation functions that contain 
$\phi_{1-}$, $\phi_{3-}$, and $\chi_{1+}$ fields 
decay exponentially. The renormalization procedure will 
affect the velocities and the critical exponents 
for the gapless fields, $\chi_{2\pm}$, $\chi_{3\pm}$, and $\phi_{2\pm}$, 
so that we should end up with renormalized 
$v^*$'s and $K^*$'s. 

In principle, the numerical values of renormalized $v^*$'s and $K^*$'s  
for finite $g$ may be obtained
from the renormalization equations 
as has been attempted for a double chain by Balents and Fisher
\cite{Balents}, although it would be difficult in practice.
However, at least in the weak-coupling limit, $g\rightarrow 0$, 
to which our treatment is meant to fall upon, 
we will certainly have $v^{*}\simeq v_F$ and $K^* \simeq 1$
for {\it gapless} modes even after the renormalization procedure.

Now we are in position 
to calculate the correlation functions. 
The two-particle correlation functions which include the following
two particle operators in the band description show power-law decay:

\noindent (1) operators constructed from two operators only in the second band
 (since the charge and spin phases are both gapless, electrons 
in this band should have the usual Luttinger liquid behavior),

\noindent (2) order parameters of singlet superconductivity in the first or 
third bands, $\psi_{1+\uparrow(\downarrow)}\psi_{1-\downarrow(\uparrow)}$,
$\psi_{3+\uparrow(\downarrow)}\psi_{3-\downarrow(\uparrow)}$. 

\noindent As a result, the order parameters that possess power-law decays 
should be the following, where we also give the exponents:

\noindent (A) The correlations within each of the two edge 
($\alpha$ and $\gamma$) chains or across the two edge chains: 

\noindent (a) $2k_F$ CDW, 
$O_{\rm intra CDW}=\psi_{\alpha(\gamma)+\uparrow}^{\dagger}
               \psi_{\alpha(\gamma)-\uparrow}$; 
$O_{\rm inter CDW}=\psi_{\alpha(\gamma)+\uparrow}^{\dagger}
               \psi_{\gamma(\alpha)-\uparrow}$,

\noindent (b) $2k_F$ SDW, 
$O_{\rm intra SDW}=\psi_{\alpha(\gamma)+\uparrow}^{\dagger}
               \psi_{\alpha(\gamma)-\downarrow}$; 
$O_{\rm inter SDW}=\psi_{\alpha(\gamma)+\uparrow}^{\dagger}
               \psi_{\gamma(\alpha)-\downarrow}$,
               
\noindent (c) singlet pairing (SS), 
$O_{\rm intra SS}=\psi_{\alpha(\gamma)+\uparrow}
               \psi_{\alpha(\gamma)-\downarrow}$; 
$O_{\rm inter SS}=\psi_{\alpha(\gamma)+\uparrow}
               \psi_{\gamma(\alpha)-\downarrow}$,
               
\noindent (d) triplet pairing (TS), 
$O_{\rm intra TS}=\psi_{\alpha(\gamma)+\uparrow}
               \psi_{\alpha(\gamma)-\uparrow}$; 
$O_{\rm inter TS}=\psi_{\alpha(\gamma)+\uparrow}
               \psi_{\gamma(\alpha)-\uparrow}$,

\noindent (B) The singlet pairing across 
the {\it central} chain ($\beta$) and an edge chain, 
$O_{\rm central SS}=\psi_{\alpha(\gamma)+\uparrow}
               \psi_{\beta-\downarrow}$. 

In the band picture we can rewright 
$O_{\rm central SS}$ as primarily comprising 
$O_{\rm central SS} \sim \psi_{1+\uparrow}\psi_{1-\downarrow}
-\psi_{3+\uparrow}\psi_{3-\downarrow}$. 
We may thus call this paring d-wave-like in a similar sense as
in the two-chain case, in which a pair is called d-wave 
when 
the pairing within the bonding band and that within 
antibonding band enter with opposite signs.\cite{Schulz,Noack}

Thus the edge-chain SDW correlation has a power-law decay, 
while the SDW correlation within the central chain 
decays exponentially since it consists of 
the terms containing $\phi_{1-}$ and/or $\phi_{3-}$ phases.  
Although we calculate the case away from half-filling, 
the SDW correlation should obviously be 
more enhanced at half-filling. 
NMR experiments at half-filling\cite{kitaoka} show 
that the nuclear-spin relaxation rate $1/T_1$ 
which is represented by the imaginary part 
of the dynamical susceptibility 
increase with decreasing temperature 
for the three-chain cuprates 
in contrast to the two-chain case. 
This is consistent with the present result, 
since the experiments should detect the total 
SDW correlation of all the chains. 

Intra- or inter-edge correlation functions 
have to involve forms bilinear in $c_2$ in eq.(3).  
They are described in terms of the second band 
$\theta_2$, which does not contain $\chi_1$ (eq.(6)), 
a phase-fixed field.  Thus the edge-channel correlations are 
completely determined by the character of the second band 
(the Luttinger-liquid band), while the other phase fields, being gapful, 
are irrelevant. 
The final result for the edge-channel correlations 
at large distances, up to $2k_F$ oscillations, is 
as follows regardless of whether the correlation is intra- or inter-edge: 
\begin{eqnarray}
\langle O_{\rm CDW}(x)O_{\rm CDW}^{\dagger}(0)\rangle&\sim&
x^{-\frac{1}{3}(K_{\rho 2}^*+2K_{\rho 3}^*)-K_{\sigma 2}^*}, 
\nonumber\\
\langle O_{\rm SDW}(x)O_{\rm SDW}^{\dagger}(0)\rangle&\sim&
x^{-\frac{1}{3}(K_{\rho 2}^*+2K_{\rho 3}^*)-\frac{1}{K_{\sigma 2}^*}}, 
\nonumber\\
\langle O_{\rm SS}(x)O_{\rm SS}^{\dagger}(0)\rangle&\sim&
x^{-\frac{1}{3}(\frac{1}{K_{\rho 2}^*}+\frac{2}{K_{\rho 3}^*})
-K_{\sigma 2}^*}, \\
\langle O_{\rm TS}(x)O_{\rm TS}^{\dagger}(0)\rangle&\sim&
x^{-\frac{1}{3}(\frac{1}{K_{\rho 2}^*}+\frac{2}{K_{\rho 3}^*})
-\frac{1}{K_{\sigma 2}^*}}. 
\nonumber
\end{eqnarray}

By contrast, if we look at the pairing $O_{\rm central SS}(x)$ across the 
central chain and one of the edge chains, 
this pairing, which circumvents the on-site repulsion 
and is linked by the resonating valence bonding between 
the neighboring chains, 
is expected to be stronger than other correlations as in the two-chain case. 
The correlation function for $O_{\rm central SS}(x)$ 
is indeed calculated to be 
\begin{eqnarray}
\langle O_{\rm central SS}(x)O_{\rm central SS}^{\dagger}(0) \rangle
\sim x^{-\frac{1}{3}(\frac{1}{K_{\rho 2}^*}+\frac{1}{2K_{\rho 3}^*})},
\end{eqnarray}

In the weak (infinitesimal) interaction limit, all the 
$K^*$'s tend to unity, where the SS exponent becomes 
as small as 
1/2 while the exponents of other correlations tend to 2.  
Thus, at least in this limit, the central SS correlation dominates over
the others. 
The duality relation (in which the pairing and density-wave 
exponents are reciprocal of each other\cite{Nagaosa}) 
is similar to that in the two-chain case, in which the 
interchain-SS exponent is 1/2 while the exponent of the $4k_F$ CDW is 2. 

In summary, we have studied correlation functions 
using the bosonization method at the renormalization-group 
fixed point away from half-filling. 
We found that the dominant correlation 
is the interchain singlet pairing across the 
central chain and either of the edge chains. 
The key message is that there is an example where the 
dominance of superconductivity only requires the 
existence of gap(s) in some spin mode, despite the 
coexistence of a power-law spin-spin correlation, 
when there are multiple modes.
It would be interesting to further look into how the situation 
for the single, double, triple, ..., chains crosses over to the 
two-dimensional system.

We wish to thank E. Arrigoni for sending us his 
work prior to publication.


\begin{references}
\bibitem{vf} The difference in fermi velocities of three bands
will not be important qualitatively as long as we consider
the case where three bands cross the fermi energy, 
for which Arrigoni's result falls on the same strong coupling fixed point 
on the plane of inter-chain hopping and filling.
\bibitem{majo} The signs assigned to cosine terms 
depend on the definition of the product of majorana 
fermion operators (here we choose 
$\eta_{i+\uparrow}\eta_{i-\downarrow}
=\eta_{i+\downarrow}\eta_{i-\uparrow}=i$), 
but this does not affect the correlation functions.
\bibitem{mathe} Mathematically, the reason why the exponents 
for the edge-channel SS correlations have to be 
greater than that for $O_{\rm central SS}$ 
is that $O_{\rm SS}$ contains a summation 
$\psi_{1+\uparrow}\psi_{1-\downarrow} + 
\psi_{3+\uparrow}\psi_{3-\downarrow}$, 
which is proportional to ${\rm cos}(\frac{\chi_{1-}}{\sqrt{2}})\sim 0$, 
while $O_{\rm central SS}$ is the subtraction, 
$\psi_{1+\uparrow}\psi_{1-\downarrow} - 
\psi_{3+\uparrow}\psi_{3-\downarrow}$, 
which is proportional to ${\rm sin}
(\frac{\chi_{1-}}{\sqrt{2}})\sim 1$.  
This is the same mechanism by which the inter-chain SS 
becomes dominant (while the intra-chain SS decays exponentially) 
in a two-chain system.
\end{references}
\end{document}